\documentclass{article}
\usepackage{fullpage}
\usepackage{amsthm}
\usepackage{amsmath}
\usepackage{amsfonts}
\usepackage{amssymb}
\usepackage{url}
\usepackage{algorithm}
\usepackage{algpseudocode}
\usepackage{easy-todo}
\usepackage{color}
\usepackage{afterpage}
\usepackage{subcaption}
\usepackage{graphicx}
\usepackage{epstopdf}
\usepackage{lineno}
\usepackage[all]{xy}
\usepackage[numbers]{natbib}  
\usepackage{setspace}
\usepackage{fullpage}
\usepackage{rotating}
\usepackage{longtable}
\usepackage{framed}
\usepackage{placeins}

\definecolor{lightgray}{gray}{0.95}

\allowdisplaybreaks

\newcommand{\authornote}[1]{}

\newcommand{\anonymiseForJETS}[1]{***}
\newcommand{\ignore}[1]{}

\newcommand{\tally}{t}

\newcommand{\proj}{p}

\newtheorem{theorem}{Example}
\newenvironment{example}{\begin{theorem}}{\end{theorem}}

\newtheorem{mydef}{Definition}
\newenvironment{definition}{\begin{mydef}\rm}{\end{mydef}}

\newenvironment{ctabbing}%
    {\begin{center}\begin{minipage}{\textwidth}\begin{tabbing}}%
    {\end{tabbing}\end{minipage}\end{center}}
					
\begin{document}

\newcommand{\cand}{\mathcal{C}}
\newcommand{\altcand}{\mathcal{A}}
\newcommand{\gone}{\mathcal{E}}
\newcommand{\stand}{\mathcal{S}}
\newcommand{\election}{\mathcal{B}}
\newcommand{\voters}{V}

\newcommand{\plusplus}{+\!\!+}
\newcommand{\mss}{\{\!\!\{}
\newcommand{\mse}{\}\!\!\}}


\title{Computing the Margin of Victory in Preferential Parliamentary Elections}
\author{Michelle Blom\footnote{Corresponding author.}, Peter J. Stuckey, Vanessa J. Teague \\
\{michelle.blom, vjteague, pstuckey\}@unimelb.edu.au \\
\doublespacing Department of Computing and Information Systems \\The University of Melbourne\\ Parkville, Australia}

\date{}
\maketitle

\begin{abstract}
We show how to use automated computation of election margins to assess the number of votes that would need to change in order to alter a parliamentary outcome for single-member preferential electorates.  In the context of increasing automation of Australian electoral processes, and accusations of deliberate interference in elections in Europe and the USA, this work forms the basis of a rigorous statistical audit of the parliamentary election outcome.  Our example is the New South Wales Legislative Council election of 2015, but the same process could be used for any similar parliament for which data was available, such as the Australian House of Representatives given the proposed automatic scanning of ballots.
\end{abstract}

\section{Introduction}\label{sec:Intro}
It is obvious that the party that wins a majority of seats in a parliamentary election may not be the party that wins a majority of votes.  This has been examined extensively in the United States \citep{tufte1973relationship, yang2008democracy}.  In Australian parliamentary elections, even the notion of a “popular majority” is poorly defined because Australian voters rank their candidates in order of preference.  But similar results occur: sometimes in practice the Parliamentary winner is not the popular majority winner and there are even some systematic biases~\citep{jackman1994measuring}.  Nevertheless it is often assumed by the public and the media that a party that wins a comfortable overall margin will comfortably win the parliamentary election.  Of course, this is not necessarily true.  

In this paper we focus on computing the Parliamentary election margin: the minimal number of votes that need to be changed, in a particular election outcome, to switch the Parliamentary winner.  This may be much less than the margin between the popular votes of the two major parties.

There are two ways that an Australian parliamentary election may be closer than it seems.  First, there may be many seats held by a very small margin.  Second, even within one seat, the margin may be smaller than it appears: Australia's preferential voting system proceeds by iteratively eliminating candidates until only two remain, then selecting the one with a larger total.  A naive observer might think that the margin of victory is the number of votes that need to be switched to reverse the winner in this last step ({\it i.e.} half the difference in the final tallies)---we call this the \emph{last-round margin}.  However, the true margin may be much smaller, because changing an early elimination step may cascade into a completely different elimination order and elect a different candidate.  Computing the correct margin for preferential voting is, in general, a computationally difficult problem, but we have demonstrated an efficient solution \citep{blom16}.  

In this paper we show how to automate both computations to compute the  Parliamentary election margin.  This requires a slight modification to our previous work to compute the winning margin \emph{with respect to a specific set of alternate winners} in each seat, combined with a simple process of adding the margins in the necessary number of seats.
 
As an example, we use data from the 2015 NSW state election to compute exactly the margin by which the Liberal/National coalition won.  The popular margin was high---the Liberal/National coalition won 46\% of formal first-preference votes compared with 34\% for the Labor parties and 10\% for The Greens.\footnote{This is from http://pastvtr.elections.nsw.gov.au/SGE2015/la/state/formal/index.htm}  The coalition won 54 seats compared to Labor's 34.  We find, however, that the number of votes necessary to switch the parliamentary outcome is less than 0.1\%. 

In prior work on US elections, we found that the true margin is almost always the last-round margin, though exceptions did occur.  
This is also true of the NSW 2015 election where, for example, 
the Lismore seat has a last-round margin of 1173, but the true margin of victory is only 209. 

The source code used to compute our results is at  \url{https://github.com/michelleblom/margin-irv}.

The same techniques and code could be easily applied to any parliamentary outcome for which complete vote data was available.  This sort of analysis could become standard procedure for any parliamentary election with automated ballot scanning. 

\subsection{Summary of results}

Of the 4.56 million votes cast in the NSW state election, it would take:
\begin{itemize}
\item 22,746 vote changes for the ALP/CLP to gain the 13 additional seats they need to win government (with 47 seats),
\item 16,349 vote changes for a Labor/GRN coalition to gain the 10 additional seats they need to win government, and
\item 10,398 vote changes to lose the Liberal/National coalition 8 seats and hence produce a hung parliament.
\end{itemize}

\subsection{Application to auditing and accuracy testing, including Australian federal elections}

The same program could be used, whenever data was available, to check automatically whether a known problem in an election was large enough to change the outcome.  Similarly, when a known number of votes were received over an insecure or unscrutinisable channel,  this could be used to decide whether that might have been enough to alter the outcome. 

Conversely, it could be used to generate evidence that the election outcome is right.

These calculations could be used as the basis for a rigorous risk-limiting audit  to confirm (or overturn) the announced election outcome.  Risk limiting audits \citep{lindeman2012gentle} take an iterative random sample of the paper ballots to check how well they reflect the announced outcome.  An audit has \emph{risk-limit} $\alpha$ if a mistaken outcome is guaranteed to be detected with a probability of at least $1-\alpha$.  Either the audit concludes with a certain confidence that the outcome is right, or it finds so many errors that a full manual recount is warranted.   The audit process is parameterised by the margin of victory in the election.   Kroll {\it et al.} \cite{kroll1001efficiently} have devised audits for parliamentary outcomes but, like most US research, they  focus on simple first-past-the-post elections in which the margin is obvious.

This is particularly important now that the Australian Parliament's Joint Standing Committee on Electoral Matters has recommended automated scanning of the ballot papers~\cite{JSCEMInterim3}.  The overall House of Representatives margin could be quickly calculated using our methods.  Rigorous risk-limiting audits could then be performed for each electorate, immediately after the election, in order to provide evidence that the overall election outcome was correct.

In a time where outside influencing of elections is a constant source of news, and where more and more elections systems involve electronic systems, either for voting or counting votes, it is critical that we have mechanisms in place to generate evidence of accurate election results, and indeed to check what degree of manipulation must have taken place for the election result to have been altered.  

\section{Background}
\subsection{Background on Australian election counting}

The lower houses of parliaments in the Australian federal and state elections are the result of a number of independent Instant Runoff Voting (IRV) elections for 
a set of single-member electorates (seats).
Each seat has a number of candidates, and 
each vote consists of
an ordered list of the candidates for that seat.\footnote{Most Australian elections require all preferences to be filled in, but some allow partial lists or several equal-last candidates.  Our analysis extends to all these cases.} 

The tallying of votes in an IRV election proceeds by a series of rounds in
which the candidate with the lowest number of votes is eliminated (see
Figure \ref{alg:IRV}) with the last remaining candidate declared the winner.
All votes in an eliminated candidate's  tally are distributed to the next 
most-preferred (remaining) candidate in their ranking. 

\begin{figure}[t]
\centering
\begin{framed}
\begin{ctabbing}
xx \= xx \= \kill
Initially, all candidates remain standing (are not eliminated)\\
\textbf{While} there is \textit{more than one} candidate standing \\
\> \textbf{For} every candidate $c$ standing\\
\> \> Tally (count) the votes in which $c$ is the highest-ranked \\
\>\> candidate of those standing\\
 \> Eliminate the candidate with the smallest tally\\
The winner is the one candidate not eliminated
\end{ctabbing}
\vspace{-0.5cm}
\end{framed}
\caption{An informal definition of the IRV counting algorithm.}
\label{alg:IRV}
\end{figure}

Let $\cand$ be the set of candidates in an IRV election $\election$. We refer to
sequences of candidates $\pi$ in list notation (e.g., $\pi =
[c_1,c_2,c_3,c_4]$), and use such sequences to represent both votes and
elimination orders. We will often treat a sequence as the set of elements it
contains. An election $\election$ is defined as a multiset\footnote{A multiset
allows for the inclusion of duplicate items.} of votes, each vote $b \in
\election$ a sequence of candidates in $\cand$, with no duplicates, listed in
order of preference (most preferred to least preferred). Let
$\mathit{first}(\pi)$ denote the first candidate appearing in sequence $\pi$
(e.g., $\mathit{first}([c_2,c_3]) = c_2$). In each round of vote counting, there are a current set of eliminated candidates $\gone$ and a current set of candidates still standing $\stand = \cand \setminus \gone$. The winner $c_w$ of the election is the last standing candidate.

Each candidate $c \in \cand$ has a \textit{tally} of votes. Votes are added to this tally upon the elimination of a candidate $c' \in \cand \setminus \{c\}$, and are redistributed from this tally upon the elimination of $c$. 

\begin{definition}{\textbf{Tally} $\mathbf{\tally_\stand(c)}$} Given candidates
$\stand \subseteq \cand$ are still standing in an election $\election$, the
tally for candidate $c \in \cand$, denoted $\tally_\stand(c)$, is defined as
the number of votes $b \in \election$ for which $c$ is the most-preferred
candidate of those remaining. Let $\proj_\stand(b)$ denote the sequence of candidates mentioned in $b$ that are also in $\stand$. 
\begin{equation}
\tally_\stand(c) = ~|~ [b ~|~ b \in \election, c = \textit{first}(\proj_\stand(b))] ~|~
\end{equation} 
\label{def:Tally}
\end{definition}

\begin{definition}{\textbf{Margin of Victory (MOV)}} The MOV in an election with candidates $\cand$ and winner $c_w \in \cand$, is the \textit{smallest} number of votes whose ranking must be modified (by an adversary) so that a candidate $c' \in \cand \setminus \{c_w\}$ is elected.  
\label{def:MOV}
\end{definition}

Often the last round margin (LRM) is used as a proxy for the margin of victory.

\begin{definition}{\textbf{Last Round Margin ($LRM$)}} The last round
margin of election, in which two candidates $\stand = \{c, c'\}$
remain with $\tally_{\stand}(c)$ and $\tally_{\stand}(c')$ votes in their
tallies, is equal to half the difference between the tallies of $c$ and $c'$ rounded up.
\begin{equation}
LRM = \lceil \frac{|\tally_{\stand}(c) - \tally_{\stand}(c')|}{2} \rceil
\end{equation}
\label{def:LRM}
\end{definition}

In this paper we will be interested in a more restricted version of margin of victory, which is the margin of victory over a subset of the non-winning candidates.

\begin{definition}{\textbf{Margin of Victory over Candidates $\altcand$ (MOVC)}} The MOVC in an election with candidates $\cand$ and winner $c_w \in \cand$ over the alternate candidates $\altcand \subseteq \cand \setminus \{c_w\}$, is the \textit{smallest} number of votes whose ranking must be modified (by an adversary) so that a candidate $c' \in \altcand$ is elected.  
\label{def:MOVC}
\end{definition}

While the MOV calculates the number of votes required to be changed to alter the winner, the MOVC calculates the number of votes required to alter the winner to one of a set $\altcand$. We will require this finer information 
in order to calculate the smallest number of votes for a different party or coalition to win the election.

\begin{example}
Consider an election between 3 candidates $a, b$, and $c$ with election $\{a:55, ca:25, bc:41, c:15\}$. The initial tallies for the candidates are $\{a:55,b:41,c:40\}$ hence $c$ is eliminated. The resulting votes are then
$\{a:80, b:41 \}$ and with tallies $\{a:80, b:41\}$
giving $a$ the victory with a last round margin of victory of 20.  
But consider changing 1 vote from $bc$ to $c$. Then the initial tallies are $\{a:55,b:40,c:41\}$
and $b$ is eliminated. The resulting votes are
$\{a:55, ca:25, c:56\}$. Then $c$ is the winner of the election. Clearly the MOV is 1 vote. 
The MOVC for $\{b\}$ is 10, which is achieved by changing 5 votes from $a$ to $bc$ and 5 from $a$ to $c$ giving first round tallies $\{ a:45,b:46,c:45\}$
where an adversary can choose to eliminate $a$ leaving votes
$\{ ca:25, bc:46, c:20\}$ with tallies $\{b:46,c:45\}$
and $b$ winning the election. \qed
\end{example}

\subsection{Automated margin computation for an IRV election}

\citet{blom16} present a branch-and-bound algorithm (denoted \textit{margin-irv}) for efficiently computing the margin of victory in an IRV election. This algorithm improves upon an existing method by \citet{Magrino:irv}.   Given an IRV election with winning candidate $c_w$, this algorithm traverses a tree defining all possible \textit{alternate} orders of candidate elimination (that result in a candidate other than $c_w$ winning the election). As the algorithm explores these alternate elimination sequences, it solves a linear program (LP) to determine the minimum number of vote manipulations required to realise each elimination order. The ultimate goal is to find an elimination sequence, in which an alternate winner is elected, that requires the smallest number of vote changes to realise. Searching through the entire space of alternate elimination sequences would be too combinatorially complex, however, and so \textit{margin-irv} incorporates rules for pruning sections of this tree from consideration. The result is an efficient algorithm for computing electoral margins.

Given an IRV election with candidates $\cand$ and winner $c_w \in \cand$, the \textit{margin-irv} algorithm starts by adding $|\cand| - 1$ partial elimination sequences to the search tree, one for each of alternate winner $c_w' \in \cand \setminus \{c_w\}$. These partial sequences form a frontier $F$. Each of these sequences contains a single candidate -- the alternate winner in question. Following the basic structure of a branch-and-bound algorithm, we compute, for each partial sequence $\pi \in F$, a lower bound on the number of vote changes required to realise a elimination sequence that \textit{ends} in $\pi$. These lower bounds are used to guide construction of the search tree. The partial sequence $\pi$ with the smallest lower bound is selected and \textit{expanded}. For each candidate $c \in \cand$ that is not already present in $\pi$, we create a new sequence with $c$ appended to the front. For example, given a set of candidates $c_1,$ $c_2,$ and  $c_3$, with winning candidate $c_3$, the partial sequence $\pi$ $=$ $[c_2]$ will be expanded to create two new sequences $[c_1,$ $c_2]$ and $[c_3,$ $c_2]$. We evaluate each new sequence $\pi'$ created by assigning it a lower bound on the number of votes required to realise any elimination order ending in $\pi'$. 

While exploring and building elimination sequences, \textit{margin-irv} maintains a running \textit{upper bound} on the value of the true margin. This upper bound is initialised to the last round margin of the election. When a sequence $\pi$ containing all candidates is constructed, our LP computes the exact number of vote manipulations required to realise it. If this number is lower than our current upper bound, the upper bound is revised, and all orders on our frontier with a lower bound greater than or equal to it are pruned from consideration (removed from our frontier). This process continues until our frontier is empty (we have considered or pruned all possible alternate elimination sequences). The value of the running upper bound is the true margin of victory of the election.

Its easy to extend our \textit{margin-irv} algorithm to also calculate MOVC for a set of alternate winners $\altcand$.  In the first step of the algorithm, rather than adding a node for each alternate winner in $\cand \setminus \{c_w\}$ we add then only for the alternates we are interested in $\altcand$. The remainder of the algorithm is unchanged.



\section{Calculating the Number of Votes to Change a Parliamentary Election Outcome}

Given a set $S$ of seats in a parliament, a winning coalition of parties $P$ 
is a set of parties such that the number of seats won by that coalition is at least some defined 
threshold $T$. Usually $T = \lceil \frac{|S|+1}{2} \rceil$, that is the coalition wins more than half the seats.
In the NSW Legislative Assembly there are 93 seats overall, so 47 are required to win government.

We can use this to calculate the number of vote changes required to change a parliamentary election result as follows.
Assume the coalition won $W \geq T$ seats.
We calculate the MOVC for each seat $s$ won by the coalition $P$ for the set of alternate candidates in that election of parties different than $P$. We then sort the MOVC values, and choose the $W - T + 1$ seats $O$
with the least MOVC values. The sum of the MOVC of these seats $O$ is the number of changes in votes required to remove the victory of the winning coalition $P$, and hence change the outcome of the election.  

Note that if the coalition is a single party $P = \{p\}$, or more generally if no seat has two candidates from the coalition, then the MOVC values required are identical to MOV values.
This is the case for the NSW Legislative Election where no seat has both a LIB and NAT candidate.

We can use a similar approach to calculate the number of vote changes required to change a parliamentary election outcome
so that another coalition $P'$ would win instead. Assume $P'$ won $W' < T$ seats.
We calculate the MOVC for each seat $s$ not won by coalition $P'$ for the set of alternate candidates in that seat with parties in $P'$. We then sort the MOVC values, and choose the $W' - T$ seats $O'$ 
with the least MOVC values. The sum of the MOVC of these seats $O'$ is the number of changes in votes required to give a parliamentary victory to the coalition $P'$.

Again if the coalition $P'$ was always the alternate winner in the calculation of the MOV, then the MOVC and MOV calculations will coincide, and indeed if $P'$ is a strong existing coalition it is likely that it is the alternate winner in most seats with the lowest MOV.

\section{Results}

The NSW Legislative Assembly Parliamentary Election of 2015 was contested by major parties:
Liberal (LIB), National (NAT), Green (GRE), Labor (ALP) and Country Labor (CLP); as
well as a number of minor parties and independents (IND). 

We found 5 seats in which the true margin was not the last-round margin: \\

\begin{tabular}{lllll}
Seat &  Candidates & Last-round margin & True margin & Winner \\
\hline
Lismore	& 6	& 1173 &	209 &	NAT \\
Balina &	7	& 1267 &	1130	& GRE \\
Heffron & 5 & 5835 & 5824 & ALP \\
Maitland &	6 &	5446 &	4012 &	CLP \\
Willoughby	& 6 &	10247 &	10160 &	LIB \\
\end{tabular}\\

The Liberal/National coalition won 54 seats to have a winning majority. 
In order to lose this majority, then need to lose $54 - 47 + 1 = 8$ seats. 
Since no seat ran both a Liberal and National candidate, we can use the MOV values to 
calculate the number of votes required to lose 8 seats. 
The 8 LIB/NAT seats with the lowest MOV are below.  Note that for Lismore, the MOV differs substantially from the last-round margin. \\

\begin{tabular}{lllll}
Seat &  Candidates & Last-round margin & True margin & Winner \\
\hline
East Hills & 5 & 189 & {189 } & {LIB }\\
Lismore & 6 & 1173 & {209 } &{ NAT} \\
Upper Hunter & 6 & 866 &  {866} & { NAT} \\
Monaro & 5 & 1122 & {1122} & { NAT} \\
Coogee & 5 & 1243 & {1243} & { LIB} \\
Tweed & 5 & 1291 & {1291} & { NAT} \\
Penrith & 8 & 2576 & {2576} & { LIB} \\
Holsworthy & 6 & 2902 & 2902 & LIB \\
\end{tabular}\\

Hence he total number of votes required to lose the majority is 10,398.

For the Labor and Country Labor coalition to win the election we 
need them win to $47 - 34 = 13$ more seats. 
The 13 seats with the lowest MOVC for a change to ALP/CLP are: \\

\begin{tabular}{llllll}
Seat &  Candidates & Last-round margin & True margin & Winner & MOVC (ALP/CLP)\\
\hline
East Hills & 5 & 189 & 189 & LIB & 189\\
Lismore & 6 & 1173 & 209 & NAT & 209\\
Upper Hunter & 6 & 866 & 866 & NAT & 866\\
Monaro & 5 & 1122 & 1122 & NAT & 1122\\
Balina & 7 & 1267 & 1130 & GRE & 1130\\
Coogee & 5 & 1243 & 1243 & LIB & 1243\\
Tweed & 5 & 1291 & 1291 & NAT & 1291\\
Balmain & 7 & 1731 & 1731 & GRE & 1731\\
Penrith & 8 & 2576 & 2576 & LIB & 2576\\
Holsworthy & 6 & 2902 & 2902 & LIB & 2902\\
Goulburn & 6 & 2945 & 2945 & LIB & 2945\\
Oatley & 5 & 3006 & 3006 & LIB & 3006\\
Newtown & 7 & 3536 & 3536 & GRE & 3536\\
\end{tabular}\\

The total number of votes required to give an ALP/CLP victory is hence 22,746.
In this case we can see, since the ALP/CLP is a strong alternate coalition, that all the MOVC calculations agree with the MOV calculations. 
Note that this is not true for all seats. 
For example in the NSW data Sydney is the first seat where the MOVC (= 5583) for the ALP and CLP coalition is different from the MOV (=2864).  This is because the runner-up was an Independent.
Note that if we used MOV instead of MOVC we would incorrectly treat Sydney as one of the seats to change, and incorrectly calculate the number of votes required for an ALP/CLP coalition to win.

The full results for all seats are in the Appendix.  The total numbers for changing the parliamentary outcome are computed by simply adding together the smallest margins for  the necessary number of seats.

\section{Conclusion} \vspace{-0.2cm}
We have shown an efficient method of automated margin computation that can be used to identify the minimum number of vote changes (or errors) necessary to alter a parliamentary election outcome using single-member preferential voting.
Our example was the NSW Legislative Assembly election of 2015, but the same tools and techniques could be immediately applied to any other parliament constructed in the same way for which full voting data was available, such as the Australian House of Representatives.

All the necessary code is openly available at \url{https://github.com/michelleblom/margin-irv}.

Accurate electoral margins can form the basis of rigorous statistical auditing of paper ballot records to check the official election result.  This would be valuable in any scenario, but is particularly important when an electronic (and hence unobservable) process such as automated ballot scanning is part of the count.  Since these are exactly the scenarios that tend to produce detailed vote data, this work provides the basis for a count that is automated and fast (because of automated ballot scanning) and also transparent and verifiably accurate, because of rigorous auditing given an accurately computed election margin.

\bibliographystyle{alpha}
\bibliography{STVMargins}

\appendix

\section{Full list of election margins for NSW 2015 state election}

Table \ref{tab:NSWmargins} records the last-round and true victory margins for each seat in the 2015 NSW lower house elections. In most seats, the last-round margin -- the difference between the two last candidates in the elimination order -- is the true margin. Exceptions to this rule are marked with an asterisk. The 8 Liberal/National coalition seats with the smallest margins are shown in bold.  The total of the margins of these 8 seats gives the smallest number of vote changes required to produce a hung parliament, 10,398.

\begin{center}
\begin{longtable}{lllll}
\caption{Last-round and true margins for each seat in the 2015 NSW lower house election.}
\label{tab:NSWmargins}\tabularnewline
\hline
Seat &  Candidates & Last-round margin & True margin & Winner \\
\hline
\endhead
\hline
\multicolumn{5}{c}{Continued}\\
\endfoot
\endlastfoot
Gosford & 6 & 102 & 102 & ALP \\
The Entrance & 5 & 171 & 171 & ALP \\
{\bf East Hills} & 5 & 189 & {\bf 189 } & {\bf LIB }\\
{\bf *Lismore} & 6 & 1173 & {\bf 209 } &{\bf NAT} \\
Strathfield & 5 & 770 & 770 & ALP \\
Granville & 6 & 837 & 837 & ALP \\
{\bf Upper Hunter} & 6 & 866 &  {\bf 866} & {\bf NAT} \\
{\bf Monaro } & 5 & 1122 & {\bf 1122} & {\bf NAT} \\
*Balina & 7 & 1267 & 1130 & GRE \\
{\bf Coogee} & 5 & 1243 & {\bf 1243} & {\bf LIB} \\
{\bf Tweed} & 5 & 1291 & {\bf 1291} & {\bf NAT} \\
Prospect & 5 & 1458 & 1458 & ALP \\
Balmain & 7 & 1731 & 1731 & GRE \\
Rockdale & 6 & 2004 & 2004 & ALP \\
Port Stephens & 5 & 2088 & 2088 & CLP \\
Auburn & 6 & 2265 & 2265 & ALP \\
{\bf Penrith} & 8 & 2576 & {\bf 2576} & {\bf LIB} \\
Kogarah & 6 & 2782 & 2782 & ALP \\
Sydney & 8 & 2864 & 2864 & IND \\
{\bf Holsworthy} & 6 & 2902 & {\bf 2902} & {\bf LIB} \\
Goulburn & 6 & 2945 & 2945 & LIB \\
Oatley & 5 & 3006 & 3006 & LIB \\
Campbelltown & 5 & 3096 & 3096 & ALP \\
Newcastle & 7 & 3132 & 3132 & ALP \\
Wollongong & 7 & 3367 & 3367 & ALP \\
Macquarie Fields & 7 & 3519 & 3519 & ALP \\
Newtown & 7 & 3536 & 3536 & GRE \\
Heathcote & 6 & 3560 & 3560 & LIB \\
Blue Mountains & 6 & 3614 & 3614 & ALP \\
Myall Lakes & 6 & 3627 & 3627 & NAT \\
Bega & 5 & 3663 & 3663 & LIB \\
Wyong & 7 & 3720 & 3720 & ALP \\
Londonderry & 5 & 3736 & 3736 & ALP \\
Seven Hills & 7 & 3774 & 3774 & LIB \\
Summer Hill & 7 & 3854 & 3854 & ALP \\
Kiama & 5 & 3856 & 3856 & LIB \\
*Maitland & 6 & 5446 & 4012 & CLP \\
Terrigal & 5 & 4053 & 4053 & LIB \\
South Coast & 5 & 4054 & 4054 & LIB \\
Clarence & 8 & 4069 & 4069 & NAT \\
Lake Macquarie & 7 & 4253 & 4253 & IND \\
Mulgoa & 5 & 4336 & 4336 & LIB \\
Oxley & 5 & 4591 & 4591 & NAT \\
Tamworth & 7 & 4643 & 4643 & NAT \\
Maroubra & 5 & 4717 & 4717 & ALP \\
Swansea & 8 & 4974 & 4974 & ALP \\
Ryde & 5 & 5153 & 5153 & LIB \\
Barwon & 6 & 5229 & 5229 & NAT \\
Riverstone & 5 & 5324 & 5324 & LIB \\
Wagga Wagga & 6 & 5475 & 5475 & LIB \\
Parramatta & 7 & 5509 & 5509 & LIB \\
Charlestown & 7 & 5532 & 5532 & ALP \\
Bankstown & 6 & 5542 & 5542 & ALP \\
Blacktown & 5 & 5565 & 5565 & ALP \\
Coffs Harbour & 5 & 5824 & 5824 & NAT \\
*Heffron & 5 & 5835 & 5824 & ALP \\
Albury & 5 & 5840 & 5840 & LIB \\
Miranda & 6 & 5881 & 5881 & LIB \\
Mount Druitt & 5 & 6343 & 6343 & ALP \\
Canterbury & 5 & 6610 & 6610 & ALP \\
Fairfield & 5 & 6998 & 6998 & ALP \\
Epping & 6 & 7156 & 7156 & LIB \\
Bathurst & 5 & 7267 & 7267 & NAT \\
Hawkesbury & 8 & 7311 & 7311 & LIB \\
Wollondilly & 6 & 7401 & 7401 & LIB \\
Shellharbour & 7 & 7519 & 7519 & ALP \\
Cabramatta & 5 & 7613 & 7613 & ALP \\
Lane Cove & 6 & 7740 & 7740 & LIB \\
Drummoyne & 6 & 8099 & 8099 & LIB \\
Keira & 5 & 8164 & 8164 & ALP \\
Camden & 5 & 8217 & 8217 & LIB \\
Lakemba & 5 & 8235 & 8235 & ALP \\
Liverpool & 5 & 8495 & 8495 & ALP \\
North Shore & 7 & 8517 & 8517 & NAT \\
Murray & 8 & 8574 & 8574 & NAT \\
Hornsby & 6 & 8577 & 8577 & LIB \\
Dubbo & 7 & 8680 & 8680 & NAT \\
Port Macquarie & 5 & 8715 & 8715 & NAT \\
Cessnock & 5 & 9187 & 9187 & CLP \\
Cootamundra & 5 & 9247 & 9247 & NAT \\
Wallsend & 5 & 9418 & 9418 & ALP \\
Cronulla & 5 & 9674 & 9674 & LIB \\
Vaucluse & 5 & 9783 & 9783 & LIB \\
Baulkham Hills & 5 & 10023 & 10023 & LIB \\
Orange & 5 & 10048 & 10048 & NAT \\
Ku-ring-gai & 5 & 10061 & 10061 & LIB \\
*Willoughby & 6 & 10247 & 10160 & LIB \\
Wakehurst & 6 & 10770 & 10770 & LIB \\
Manly & 5 & 10806 & 10806 & LIB \\
Pittwater & 5 & 11430 & 11430 & LIB \\
Northern Tablelands & 6 & 11969 & 11969 & LIB \\
Davidson & 5 & 12960 & 12960 & LIB \\
Castle Hill & 5 & 13160 & 13160 & LIB \\
\hline
\end{longtable}
 \end{center}
 
Table \ref{tab:NSWchangesElectALP_CLP} lists the number of vote changes necessary to elect an ALP or CLP candidate.  This is at least the true margin (from the previous table), but may be strictly more, for example if an independent candidate was the runner-up.  The rows inside the double lines are the 10 seats with the smallest changes necessary to give the labor parties 47 seats.  The combined total number of votes needed to produce this is the sum of those rows: 22746.

\begin{center}
\begin{longtable}{lllllr}
\caption{Last-round margins, true margins, and the number of vote changes required to elect an ALP or CLP candidate for each seat in the 2015 NSW lower house election.}
\label{tab:NSWchangesElectALP_CLP}\tabularnewline
\hline
	Seat &  Candidates & Last-round margin & True margin & Winner & Changes to Elect \\
	     &             &                   &             &     & ALP or CLP\\
\hline
\endhead
\hline
\multicolumn{6}{c}{Continued}\\
\endfoot
\endlastfoot
Auburn & 6 & 2265 & 2265 & ALP & 0\\
Bankstown & 6 & 5542 & 5542 & ALP & 0\\
Blacktown & 5 & 5565 & 5565 & ALP & 0\\
Blue Mountains & 6 & 3614 & 3614 & ALP & 0\\
Cabramatta & 5 & 7613 & 7613 & ALP & 0\\
Campbelltown & 5 & 3096 & 3096 & ALP & 0\\
Canterbury & 5 & 6610 & 6610 & ALP & 0\\
Cessnock & 5 & 9187 & 9187 & CLP & 0\\
Charlestown & 7 & 5532 & 5532 & ALP & 0\\
Fairfield & 5 & 6998 & 6998 & ALP & 0\\
Gosford & 6 & 102 & 102 & ALP & 0\\
Granville & 6 & 837 & 837 & ALP & 0\\
Heffron & 5 & 5835 & 5824 & ALP & 0\\
Keira & 5 & 8164 & 8164 & ALP & 0\\
Kogarah & 6 & 2782 & 2782 & ALP & 0\\
Lakemba & 5 & 8235 & 8235 & ALP & 0\\
Liverpool & 5 & 8495 & 8495 & ALP & 0\\
Londonderry & 5 & 3736 & 3736 & ALP & 0\\
Macquarie Fields & 7 & 3519 & 3519 & ALP & 0\\
Maitland & 6 & 5446 & 4012 & CLP & 0\\
Maroubra & 5 & 4717 & 4717 & ALP & 0\\
Mount Druitt & 5 & 6343 & 6343 & ALP & 0\\
Newcastle & 7 & 3132 & 3132 & ALP & 0\\
Port Stephens & 5 & 2088 & 2088 & CLP & 0\\
Prospect & 5 & 1458 & 1458 & ALP & 0\\
Rockdale & 6 & 2004 & 2004 & ALP & 0\\
Shellharbour & 7 & 7519 & 7519 & ALP & 0\\
Strathfield & 5 & 770 & 770 & ALP & 0\\
Summer Hill & 7 & 3854 & 3854 & ALP & 0\\
Swansea & 8 & 4974 & 4974 & ALP & 0\\
The Entrance & 5 & 171 & 171 & ALP & 0\\
Wallsend & 5 & 9418 & 9418 & ALP & 0\\
Wollongong & 7 & 3367 & 3367 & ALP & 0\\
Wyong & 7 & 3720 & 3720 & ALP & 0\\
\hline \hline
East Hills & 5 & 189 & 189 & LIB & 189\\
Lismore & 6 & 1173 & 209 & NAT & 209\\
Upper Hunter & 6 & 866 & 866 & NAT & 866\\
Monaro & 5 & 1122 & 1122 & NAT & 1122\\
Balina & 7 & 1267 & 1130 & GRE & 1130\\
Coogee & 5 & 1243 & 1243 & LIB & 1243\\
Tweed & 5 & 1291 & 1291 & NAT & 1291\\
Balmain & 7 & 1731 & 1731 & GRE & 1731\\
Penrith & 8 & 2576 & 2576 & LIB & 2576\\
Holsworthy & 6 & 2902 & 2902 & LIB & 2902\\
Goulburn & 6 & 2945 & 2945 & LIB & 2945\\
Oatley & 5 & 3006 & 3006 & LIB & 3006\\
Newtown & 7 & 3536 & 3536 & GRE & 3536\\
\hline \hline
Heathcote & 6 & 3560 & 3560 & LIB & 3560\\
Myall Lakes & 6 & 3627 & 3627 & NAT & 3627\\
Bega & 5 & 3663 & 3663 & LIB & 3663\\
Seven Hills & 7 & 3774 & 3774 & LIB & 3774\\
Kiama & 5 & 3856 & 3856 & LIB & 3856\\
Terrigal & 5 & 4053 & 4053 & LIB & 4053\\
South Coast & 5 & 4054 & 4054 & LIB & 4054\\
Clarence & 8 & 4069 & 4069 & NAT & 4069\\
Lake Macquarie & 7 & 4253 & 4253 & IND & 4253\\
Mulgoa & 5 & 4336 & 4336 & LIB & 4336\\
Oxley & 5 & 4591 & 4591 & NAT & 4591\\
Ryde & 5 & 5153 & 5153 & LIB & 5153\\
Barwon & 6 & 5229 & 5229 & NAT & 5229\\
Riverstone & 5 & 5324 & 5324 & LIB & 5324\\
Wagga Wagga & 6 & 5475 & 5475 & LIB & 5475\\
Parramatta & 7 & 5509 & 5509 & LIB & 5509\\
Sydney & 8 & 2864 & 2864 & IND & 5583\\
Coffs Harbour & 5 & 5824 & 5824 & NAT & 5824\\
Albury & 5 & 5840 & 5840 & LIB & 5840\\
Miranda & 6 & 5881 & 5881 & LIB & 5881\\
Epping & 6 & 7156 & 7156 & LIB & 7156\\
Bathurst & 5 & 7267 & 7267 & NAT & 7267\\
Hawkesbury & 8 & 7311 & 7311 & LIB & 7311\\
Wollondilly & 6 & 7401 & 7401 & LIB & 7401\\
Lane Cove & 6 & 7740 & 7740 & LIB & 7740\\
Drummoyne & 6 & 8099 & 8099 & LIB & 8099\\
Camden & 5 & 8217 & 8217 & LIB & 8217\\
Hornsby & 6 & 8577 & 8577 & LIB & 8577\\
Dubbo & 7 & 8680 & 8680 & NAT & 8680\\
Port Macquarie & 5 & 8715 & 8715 & NAT & 8715\\
North Shore & 7 & 8517 & 8517 & NAT & 8798\\
Cootamundra & 5 & 9247 & 9247 & NAT & 9247\\
Murray & 8 & 8574 & 8574 & NAT & 9483\\
Cronulla & 5 & 9674 & 9674 & LIB & 9674\\
Baulkham Hills & 5 & 10023 & 10023 & LIB & 10023\\
Orange & 5 & 10048 & 10048 & NAT & 10048\\
Ku-ring-gai & 5 & 10061 & 10061 & LIB & 10061\\
Willoughby & 6 & 10247 & 10160 & LIB & 10160\\
Vaucluse & 5 & 9783 & 9783 & LIB & 10581\\
Wakehurst & 6 & 10770 & 10770 & LIB & 10770\\
Tamworth & 7 & 4643 & 4643 & NAT & 11283\\
Northern Tablelands & 6 & 11969 & 11969 & LIB & 11969\\
Manly & 5 & 10806 & 10806 & LIB & 12106\\
Pittwater & 5 & 11430 & 11430 & LIB & 12181\\
Davidson & 5 & 12960 & 12960 & LIB & 13065\\
Castle Hill & 5 & 13160 & 13160 & LIB & 13160\\
\hline
\end{longtable}
\end{center}

Table \ref{tab:NSWchangesElectALP_CLP_GRN} records the margins for a Labor-Green coalition.  In this case the total number of vote changes required to produce this outcome is 16349.

\begin{center}
\begin{longtable}{lllllr}
\caption{Last-round margins, true margins, and the number of vote changes required to elect an ALP, CLP, or Greens (GRN) candidate for each seat in the 2015 NSW lower house election.}
\label{tab:NSWchangesElectALP_CLP_GRN}\tabularnewline
\hline
	Seat &  Candidates & Last-round margin & True margin & Winner & Changes to Elect \\
	     &             &                   &             &     & ALP, CLP, or GRN\\
\hline
\endhead
\hline
\multicolumn{6}{c}{Continued}\\
\endfoot
\endlastfoot
Auburn & 6 & 2265 & 2265 & ALP & 0\\
Balina & 7 & 1267 & 1130 & GRE & 0\\
Balmain & 7 & 1731 & 1731 & GRE & 0\\
Bankstown & 6 & 5542 & 5542 & ALP & 0\\
Blacktown & 5 & 5565 & 5565 & ALP & 0\\
Blue Mountains & 6 & 3614 & 3614 & ALP & 0\\
Cabramatta & 5 & 7613 & 7613 & ALP & 0\\
Campbelltown & 5 & 3096 & 3096 & ALP & 0\\
Canterbury & 5 & 6610 & 6610 & ALP & 0\\
Cessnock & 5 & 9187 & 9187 & CLP & 0\\
Charlestown & 7 & 5532 & 5532 & ALP & 0\\
Fairfield & 5 & 6998 & 6998 & ALP & 0\\
Gosford & 6 & 102 & 102 & ALP & 0\\
Granville & 6 & 837 & 837 & ALP & 0\\
Heffron & 5 & 5835 & 5824 & ALP & 0\\
Keira & 5 & 8164 & 8164 & ALP & 0\\
Kogarah & 6 & 2782 & 2782 & ALP & 0\\
Lakemba & 5 & 8235 & 8235 & ALP & 0\\
Liverpool & 5 & 8495 & 8495 & ALP & 0\\
Londonderry & 5 & 3736 & 3736 & ALP & 0\\
Macquarie Fields & 7 & 3519 & 3519 & ALP & 0\\
Maitland & 6 & 5446 & 4012 & CLP & 0\\
Maroubra & 5 & 4717 & 4717 & ALP & 0\\
Mount Druitt & 5 & 6343 & 6343 & ALP & 0\\
Newcastle & 7 & 3132 & 3132 & ALP & 0\\
Newtown & 7 & 3536 & 3536 & GRE & 0\\
Port Stephens & 5 & 2088 & 2088 & CLP & 0\\
Prospect & 5 & 1458 & 1458 & ALP & 0\\
Rockdale & 6 & 2004 & 2004 & ALP & 0\\
Shellharbour & 7 & 7519 & 7519 & ALP & 0\\
Strathfield & 5 & 770 & 770 & ALP & 0\\
Summer Hill & 7 & 3854 & 3854 & ALP & 0\\
Swansea & 8 & 4974 & 4974 & ALP & 0\\
The Entrance & 5 & 171 & 171 & ALP & 0\\
Wallsend & 5 & 9418 & 9418 & ALP & 0\\
Wollongong & 7 & 3367 & 3367 & ALP & 0\\
Wyong & 7 & 3720 & 3720 & ALP & 0\\
\hline \hline
East Hills & 5 & 189 & 189 & LIB & 189\\
Lismore & 6 & 1173 & 209 & NAT & 209\\
Upper Hunter & 6 & 866 & 866 & NAT & 866\\
Monaro & 5 & 1122 & 1122 & NAT & 1122\\
Coogee & 5 & 1243 & 1243 & LIB & 1243\\
Tweed & 5 & 1291 & 1291 & NAT & 1291\\
Penrith & 8 & 2576 & 2576 & LIB & 2576\\
Holsworthy & 6 & 2902 & 2902 & LIB & 2902\\
Goulburn & 6 & 2945 & 2945 & LIB & 2945\\
Oatley & 5 & 3006 & 3006 & LIB & 3006\\
\hline \hline
Heathcote & 6 & 3560 & 3560 & LIB & 3560\\
Myall Lakes & 6 & 3627 & 3627 & NAT & 3627\\
Bega & 5 & 3663 & 3663 & LIB & 3663\\
Seven Hills & 7 & 3774 & 3774 & LIB & 3774\\
Kiama & 5 & 3856 & 3856 & LIB & 3856\\
Terrigal & 5 & 4053 & 4053 & LIB & 4053\\
South Coast & 5 & 4054 & 4054 & LIB & 4054\\
Clarence & 8 & 4069 & 4069 & NAT & 4069\\
Lake Macquarie & 7 & 4253 & 4253 & IND & 4253\\
Mulgoa & 5 & 4336 & 4336 & LIB & 4336\\
Oxley & 5 & 4591 & 4591 & NAT & 4591\\
Ryde & 5 & 5153 & 5153 & LIB & 5153\\
Barwon & 6 & 5229 & 5229 & NAT & 5229\\
Riverstone & 5 & 5324 & 5324 & LIB & 5324\\
Wagga Wagga & 6 & 5475 & 5475 & LIB & 5475\\
Parramatta & 7 & 5509 & 5509 & LIB & 5509\\
Sydney & 8 & 2864 & 2864 & IND & 5583\\
Coffs Harbour & 5 & 5824 & 5824 & NAT & 5824\\
Albury & 5 & 5840 & 5840 & LIB & 5840\\
Miranda & 6 & 5881 & 5881 & LIB & 5881\\
Epping & 6 & 7156 & 7156 & LIB & 7156\\
Bathurst & 5 & 7267 & 7267 & NAT & 7267\\
Hawkesbury & 8 & 7311 & 7311 & LIB & 7311\\
Wollondilly & 6 & 7401 & 7401 & LIB & 7401\\
Lane Cove & 6 & 7740 & 7740 & LIB & 7740\\
Drummoyne & 6 & 8099 & 8099 & LIB & 8099\\
Camden & 5 & 8217 & 8217 & LIB & 8217\\
North Shore & 7 & 8517 & 8517 & NAT & 8517\\
Hornsby & 6 & 8577 & 8577 & LIB & 8577\\
Dubbo & 7 & 8680 & 8680 & NAT & 8680\\
Port Macquarie & 5 & 8715 & 8715 & NAT & 8715\\
Cootamundra & 5 & 9247 & 9247 & NAT & 9247\\
Murray & 8 & 8574 & 8574 & NAT & 9483\\
Cronulla & 5 & 9674 & 9674 & LIB & 9674\\
Vaucluse & 5 & 9783 & 9783 & LIB & 9783\\
Baulkham Hills & 5 & 10023 & 10023 & LIB & 10023\\
Orange & 5 & 10048 & 10048 & NAT & 10048\\
Ku-ring-gai & 5 & 10061 & 10061 & LIB & 10061\\
Willoughby & 6 & 10247 & 10160 & LIB & 10160\\
Wakehurst & 6 & 10770 & 10770 & LIB & 10770\\
Manly & 5 & 10806 & 10806 & LIB & 10806\\
Tamworth & 7 & 4643 & 4643 & NAT & 11283\\
Pittwater & 5 & 11430 & 11430 & LIB & 11430\\
Northern Tablelands & 6 & 11969 & 11969 & LIB & 11969\\
Davidson & 5 & 12960 & 12960 & LIB & 12960\\
Castle Hill & 5 & 13160 & 13160 & LIB & 13160\\
\hline
\end{longtable}
\end{center}

\end{document}